\documentclass[final,5p,times,twocolumn]{elsarticle}
\usepackage{lineno,hyperref}
\modulolinenumbers[5]

\journal{Nuclear Instruments and Methods in Physics Research A}

\begin{document}

\begin{frontmatter}

\title{Aging suppression in Multistrip Multigap Resistive Plate Chambers for high counting rate experiments}

\author[]{ M.~Petri\c{s}\corref{mycorrespondingauthor}}
\cortext[mycorrespondingauthor]{Corresponding author}
\ead{mpetris@nipne.ro}
\author[]{V.~Aprodu}
\author[]{D.~Barto\c{s}}
\author[]{D.~Doroban\c{t}u}
\author[]{V.~Du\c{t}\u{a}}
\author[]{M.~Petrovici}
\author[]{A.~Radu}

\address[1]{Hadron Physics Department, National Institute for Physics and Nuclear Engineering (IFIN-HH), Bucharest, Romania}

\begin{abstract}
A long term operation of Multi-Strip Multi-Gap Resistive Plate Chambers (MSMGRPC) with gas mixtures based on C$_2$H$_2$F$_4$ and SF$_6$ leads to aging effects, observed as depositions on the surface of the resistive electrodes.  Moreover, enhanced depositions and higher noise rates were evidenced around the nylon spacers used for defining the gas gaps between the resistive electrodes. 
The aging effects are reflected in an increase of the dark current and dark counting rate, with negative impact on the long term performance of the chamber and data volume in a free running readout  mode operation.
MSMGRPC prototypes designed with a direct gas flow through the gas gaps and  minimization of the number of spacers in the active area were developed as mitigation solution.  Prototypes with this new design and different granularities were assembled using fishing line as spacers and investigated for aging effects.  Although a significant reduction in the dark current and dark counting rate was evidenced,  dark counting rate localized  around the fishing line spacers remains. In this paper, a new generation of direct flow chambers based on discrete spacers is presented. The results of their aging investigations show that, even at lower gas flows, the aging effects become negligible. 
\end{abstract}

\begin{keyword}
\texttt{Gaseous detectors \sep Resistive plate chamber \sep Aging effect \sep High intensity X-ray flux }
\end{keyword}

\end{frontmatter}


\section{Introduction}
\label{1}
The upgrade of the running high-energy physics experiments or construction of future ones, like Compressed Baryonic Matter (CBM) at the future FAIR acceleration facility in Darmstadt \citep{epja2017},  requires experimental setups able to cope with unprecedented interaction rates of up to  10$^7$ Hz, with the aim to investigate very rare probes with high precision and enough statistics. For charged particle identification, the CBM experimental setup contains a Time of Flight (TOF) subsystem \cite{toftdr} based on multi-gap resistive plate chambers \cite{mrpc} with multistrip readout \cite{msmgrpc1}. Due to the high interaction rates,  the chambers of the CBM-TOF wall have to cope with anticipated counting rates ranging between 0.1~kHz/cm$^2$ up to a few tens of ~kHz/cm$^2$, depending on their location \cite{toftdr, jinst2019}.  
Our R\&D activity addresses the most challenging region of the CBM~-~TOF wall \cite{jinst2014}, positioned at low polar angles between 2.5$^0$and 12$^0$, called  inner wall. The chambers of the CBM-TOF inner wall are required to maintain their performance (efficiency above 90\% and system time resolution better than 80~ps) at foreseen counting rates between 6~kHz/cm$^2$ up to 50 kHz/cm$^2$,  with an occupancy less than 5\%.

In the current design, the CBM-TOF inner wall is 
 equipped with a total number of 316 Multi-Strip Multi-Gap Resistive Plate Chambers (MSMGRPCs). As a function of polar angle, different regions of granularity are covered by three types of chambers,  differentiated only by their strip length (MRPC1a of 56~mm, MRPC1b of 96~mm and MRPC1c of 196~mm) while the pitch (9.02~mm) and the inner architecture are the same.  
In view of their use  over the lifetime of the experiment, especially in the challenging counting rates mentioned above, detailed aging investigations were performed \cite{nima2022} using a high activity $^{60}$Co source \cite{irasm}.
 A gas pollution effect was evidenced by the chemical composition of the layers  deposited on the resistive electrode surfaces  of a prototype with classical gas exchange via diffusion.  Moreover, enhanced depositions and larger noise rates were observed around the nylon spacers (fishing line) which define the gas gaps between resistive electrodes. 
On a long term operation, the aging effects are reflected in an increase of the dark current and dark counting rate which might impact the chamber performance and also lead to an artificial increase of the data volume in a free-running readout mode.
With the aim to reduce the observed aging effects, we focused our efforts to a faster removal  from the gas gaps of the chemical compounds produced during irradiation, preventing their depositions on the glass surfaces. 
We developed MSMGRPC prototypes with a direct gas flow architecture and  minimization of the number of spacers in the active area. The constructive details of the first prototypes with a gas flow through the gaps and a high granularity (MRPC1a, 56~mm strip length), together with their in-beam tests are comprehensively described in \cite{nimpisa2022}. Detailed aging investigations using high X-ray fluxes showed better behaviour of the direct flow prototype in comparison with a prototype with classical gas exchange via diffusion in terms of the dark current and dark counting rate \cite{nimrpc2022}. 
In section~\ref{2}  we present the constructive details and the results of the aging tests performed using high intensity X-ray fluxes, for the direct flow  MRPC1b and MRPC1c prototypes.  After the aging tests, their performance in terms of time resolution and efficiency were investigated using cosmic~-~rays. Furthermore, a scan of the MSMGRPC active areas in a self-triggered mode operation of the data acquisition (DAQ) system revealed still higher noisy signals localized in the close vicinity of the fishing line spacers. In order to further minimize these effects, a new generation of direct flow chambers based on polyimide  discrete spacers is proposed.  Section~\ref{3} of the paper presents the results of their aging investigations using the high intensity X-ray fluxes at reduced gas flows.
Conclusions are presented in Section~\ref{4}. 
\section{Direct flow MSMGRPC based on fishing line spacers} 
\label{2}
The typical architecture of the MSMGRPC prototypes \cite{rjp2021} is a double stack structure with 5 gas gaps per stack  of 200~$\mu$m. The resistive electrodes are made of low resistivity glass ($\sim$10$^{10}\Omega$cm)~\cite{nima2013} in order to cope with the  high counting rates.
They have a unique architecture with a strip structure for both high voltage and readout electrodes \cite{jinst2012, nima2019}, matching the signal transmission line impedance to the input impedance of the front-end electronics (FEE) \cite{rjp2018}.  
Each chamber has 32 strips distributed  with a 9.02~mm pitch along the 300~mm length  of the glass electrodes. The nylon fishing line spacers are stretched along the resistive electrode, across the strips. 
\subsection{Constructive details of the direct flow prototypes of 96~mm and 196~mm strip length}
\label{contr}
\begin{figure}[htb]
\centering
\includegraphics*[width=55mm]{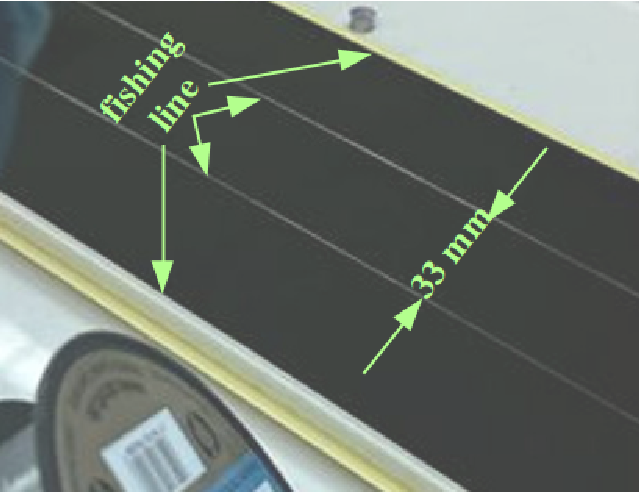}
\caption{ Two spacers are positioned inside the active area, at 33~mm distance one to the other, the two outermost spacers are positioned at 1.5~mm distance from the edges, out of the electric field.}
\label{f1a}
\end{figure}

\begin{figure}[htb]
\centering
\includegraphics*[width=50mm]{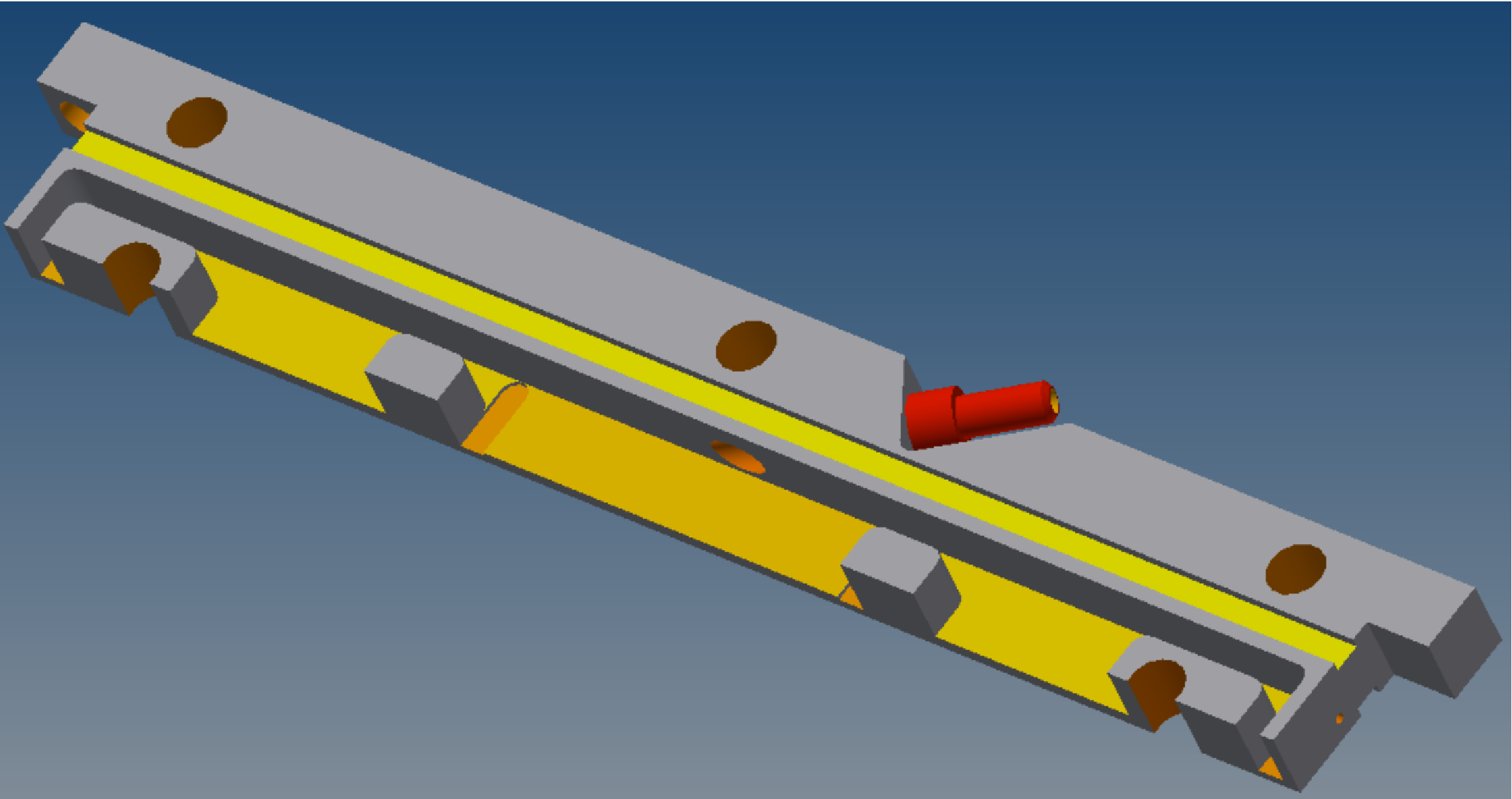}
\caption{The design of the gas guide for MRPC1b prototype. }
\label{f1b}
\end{figure}

 The length of the strips of both high voltage and read-out electrodes of MRPC1b chamber is 96~mm, distributed parallel to the 100~mm side of the resistive electrodes and centered relative to the glass edges. This allows to position two spacers out of four, outside of the active area, at 1.5~mm distance to the corresponding edge of the resistive electrode. The other two are positioned inside the active area, at 33 mm distance one to the other, as can be seen in Fig.~\ref{f1a}.  The active area is defined by the strip length (9.6~cm) multiplied by the strip pitch (0.902~cm) and the number of strips (32).
 At the two ends of each stack (short side) were positioned gas guides which force the gas  flow through the gaps between the resistive electrodes. The design of a gas guide for the MRPC1b prototype is shown in Fig.~\ref{f1b}. It has a gas pipe (red pipe in the figure) for input or output of the gas mixture on the outer side and four pillars around which the fishing line is stretched on the inner side.  Around each stack, at 1~mm distance and parallel with the long side of the resistive electrodes, are positioned two epoxy glass fiber ledges. The gas guides and the two ledges are closing the gas volume of each stack (as for MRPC1a, \cite{nimrpc2022}). 

The MRPC1c prototype has six spacers which  are distributed across the surface of the glass plate of 200~mm width, with an equal distance one to the other. The outermost ones are positioned as for MRPC1b, at 1.5~mm distance from the edge of the glass electrode, outside the active area (19.6~cm$\times$0.902~cm$\times$32), while the other four spacers are positioned inside the active area.   The gas volume of each stack was tightened in the same way as for MRPC1a and MRPC1b. 
After their assembling, the measured gas transmission through the gaps was $\sim$ 100\% for both MRPC1b and MRPC1c prototypes. 

Each chamber was closed in a gas tight and electromagnetically screened housing box and a gas mixture of 97.5\% C$_2$H$_2$F$_4$ and 2.5\% SF$_6$ was flushed through the gas gaps. The housing box was also flushed with gas mixture  at 2~l/h flow rate, in order to avoid an eventual contamination of the gas inside of the detector. The measured oxygen contamination of the gas mixture inside the gaps of the chambers was very low, $<$10~ppm. 
The applied high voltage was of 2~$\times$~6~kV for each chamber, for all the performed tests. 
The signals are readout in a differential mode, both the positive and negative picked-up signals being fed into the input of a FEE readout channel. 
\subsection{Aging tests}
\label{aging}
In a first step, the prototypes were investigated for the aging effects. 
Two mini X-ray tubes \cite{miniX} were aligned next to each other and positioned in front of the housing box, at a certain distance which assured an almost uniform exposure of the chamber to the high X-ray flux. 

The 32 strips of each MSMGRPC prototype were readout at both ends by eight front-end electronics cards based on NINO ASIC   \cite{nino}. Each card provides besides the information of the time of the rising and falling edge  for each of its eight channels, a common logic OR signal. The logic AND
between the OR signals of the two ends of each group of eight strips was fed into the input of a scaler, recording the rate of the signals.  The average counting rate per unit area of the chamber was obtained as the arithmetic mean of the recorded rates divided by the corresponding strip area. 

The MRPC1b prototype was exposed in successive days for a period of about 6~h per day to a maximum X-ray flux which corresponds to a counting rate of 23~kHz/cm$^2$. The current (provided by the power supply) and the counting rate  of the chamber were monitored during the exposure.  After the end of each period of 6~h of irradiation, the recovery of the dark current and the dark counting rate were recorded for about 2 hours. For a flow of 2~l/h, an increase of the dark current and dark counting rate with the accumulated charge was observed, similar with the results obtained for the  MRPC1a prototype \cite{nimrpc2022}, tested in similar conditions. Further on, the chamber was flushed in successive days with increasing gas flow rates (4~l/h and 6~l/h).  The obtained results confirm the ones previously obtained for prototype MRPC1a: the dark current and dark counting rate are smaller and also recover faster with the increase of the gas flow rates. 
\begin{figure}[htb]
\centering
\includegraphics*[width=43mm]{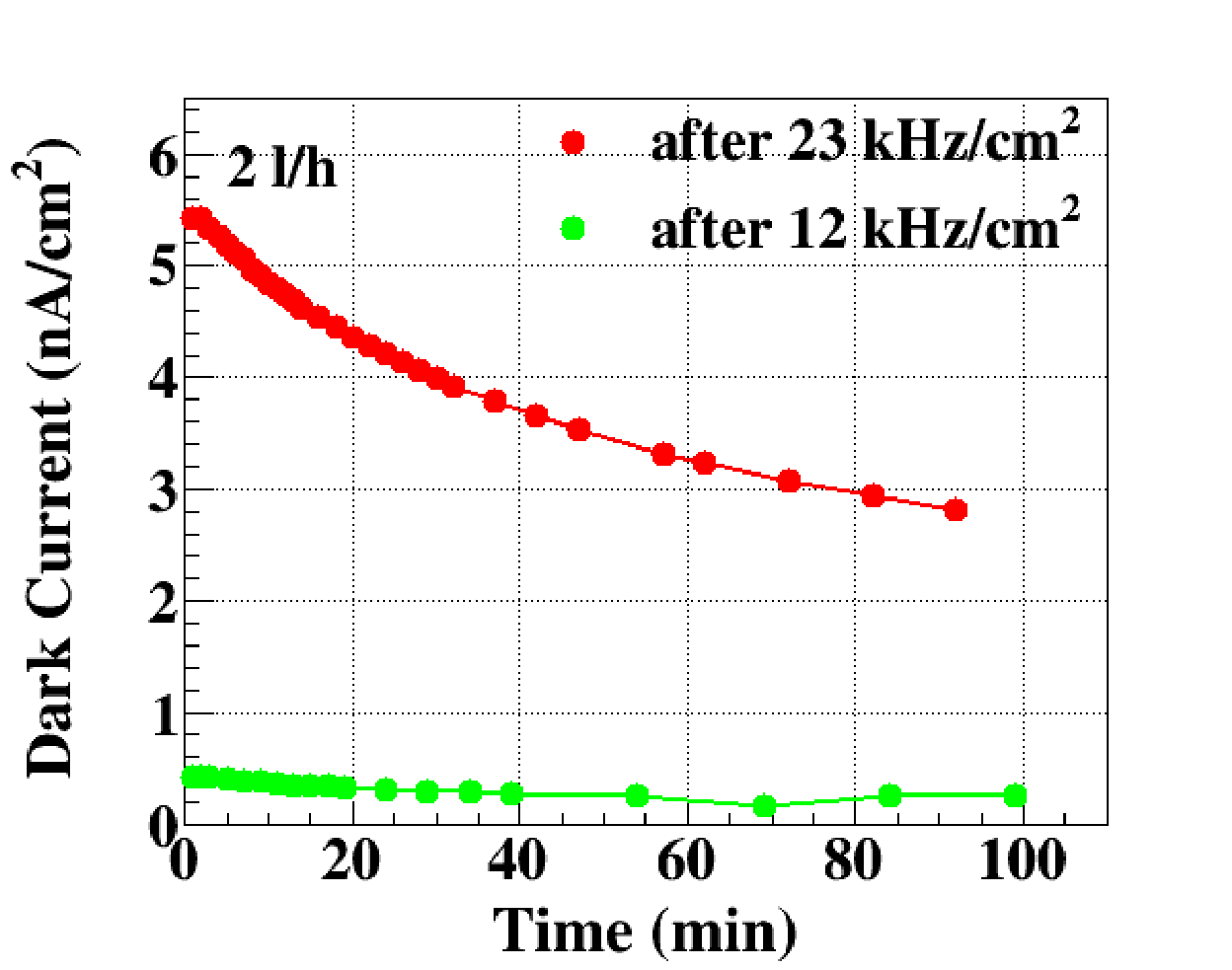}
\includegraphics*[width=43mm]{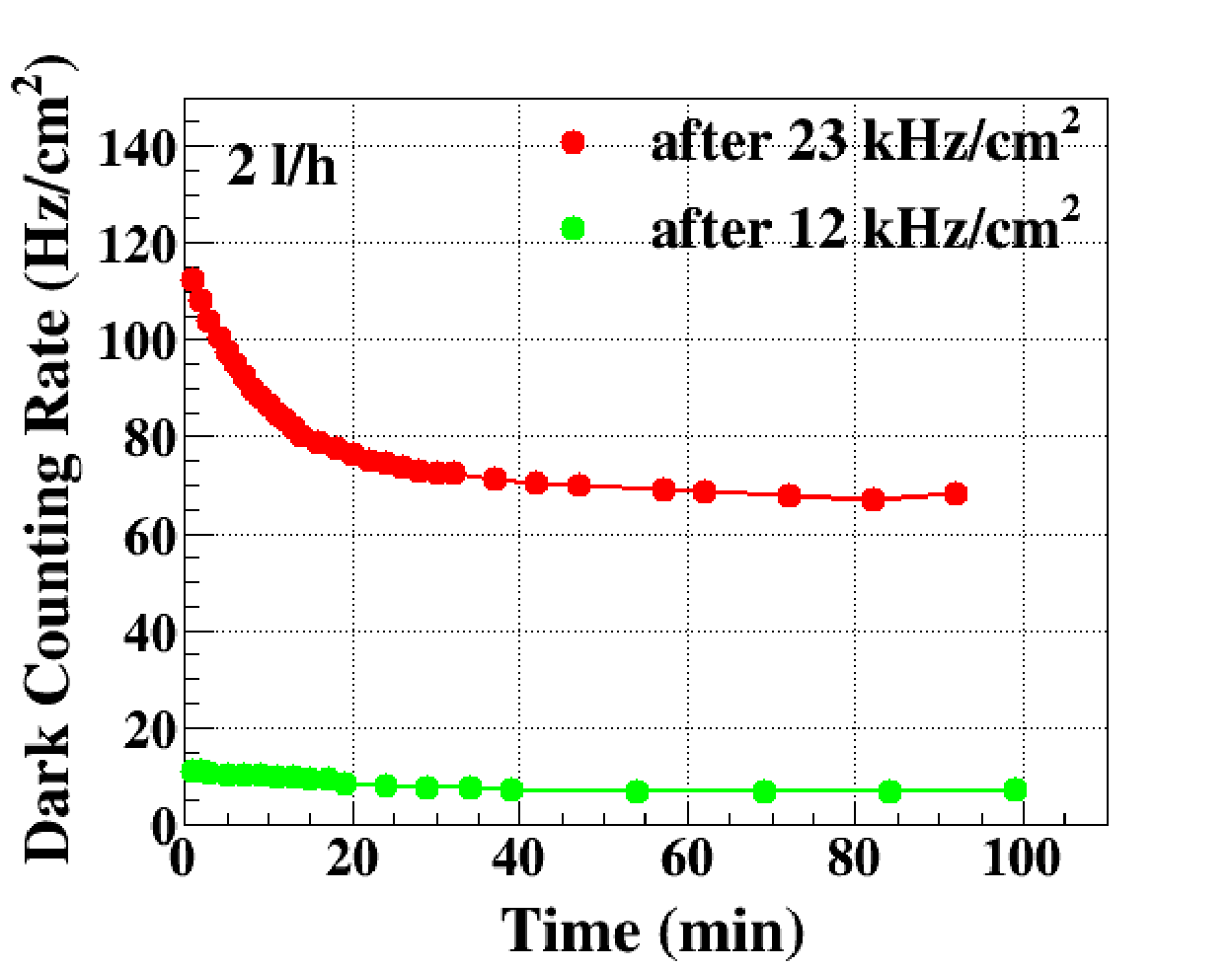}
\caption{Dark current and dark counting rate of MRPC1b after the exposure to different counting rate, at 2~l/h gas flow rate. }
\label{f3}
\end{figure}
However, taking into consideration the polar angles covered by the MRPC1b chambers in the CBM-TOF inner zone, the maximum anticipated counting rate to which they will be exposed is lower than 10~kHz/cm$^2$. Consequently, the chamber was further on exposed to a lower X-ray flux, corresponding to a 12~kHz/cm$^2$ counting rate. 
As is shown in Fig.~\ref{f3}, both dark current and dark counting rate have much lower values after 6~h exposure at 12~kHz/cm$^2$ than for exposure to  23~kHz/cm$^2$, at the same gas flow of 2~l/h flow.
\begin{figure}[htb]
\centering
\includegraphics*[width=50mm]{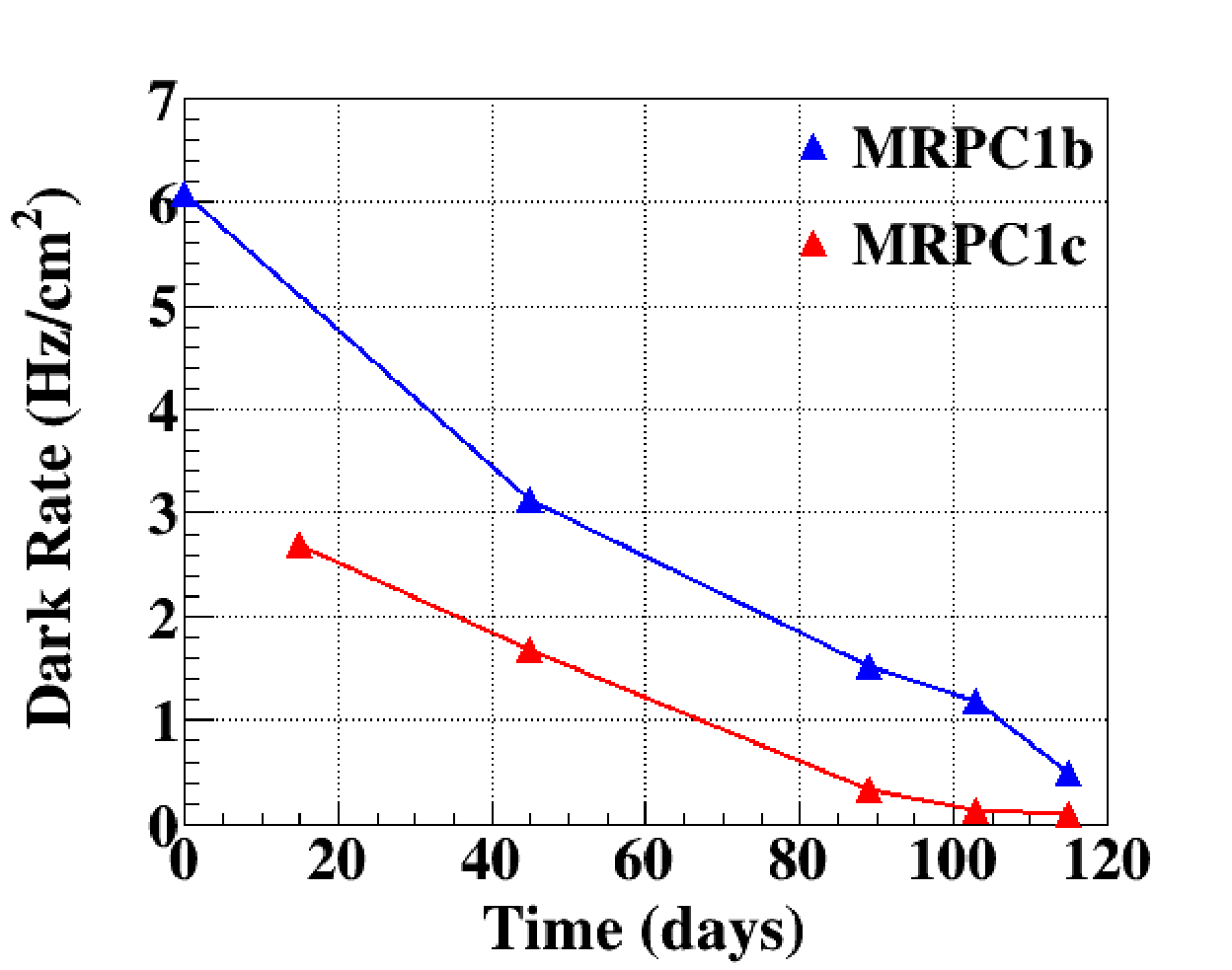}
\caption{Long term evolution of the dark counting rate. }
\label{f4}
\end{figure}

The MRPC1c chambers will be positioned in the CBM-TOF inner zone at polar angles where the counting rate  is estimated to be less than 6~kHz/cm$^2$. For this reason the MRPC1c prototype was exposed to a X-ray flux corresponding to a maximum counting rate of 8~kHz/cm$^2$. After successive exposures, the accumulated charge was of 2.4~mC/cm$^2$.  The values of the dark current (0.02~nA/cm$^2$) and dark counting rate (2.7~Hz/cm$^2$) measured 24~h after the end of the last exposure have been very low. For the MRPC1b chamber, after an accumulated charge of 9.4~mC/cm$^2$ the dark counting rate reduced to 6.1~Hz/cm$^2$ and dark current to 0.05~nA/cm$^2$ at 24~hours after the end of the last exposure. 
Moreover, they are converging in time towards low values of the dark counting rate as is shown in Fig.~\ref{f4} and also to a negligible dark current.
\subsection{Cosmic ray tests}
\begin{figure}[htb]
\centering
\includegraphics*[width=50mm]{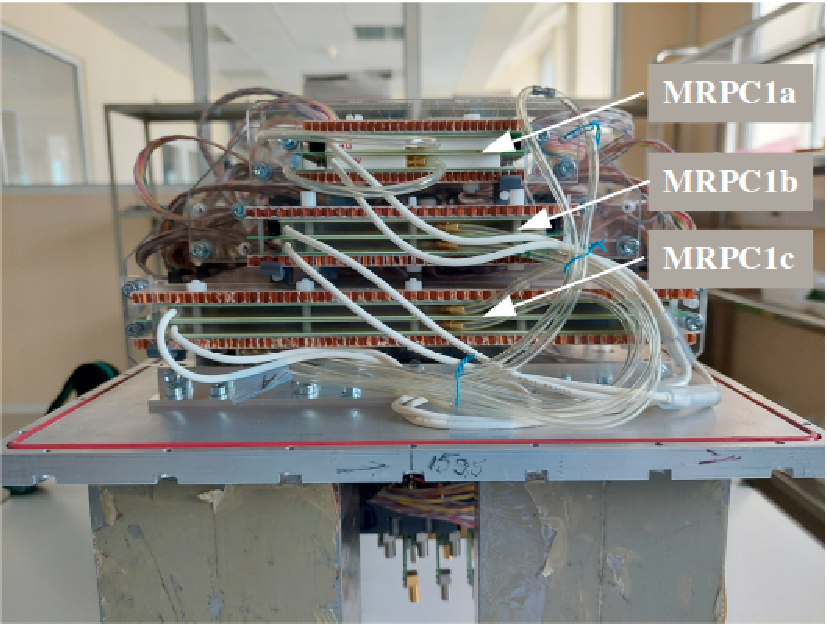}
\caption{Stacked chambers for cosmic ray test.  }
\label{f5}
\end{figure}

After the aging tests, the performance of the chambers in terms of efficiency and time resolution was tested using cosmic rays. We stacked all tree chambers, closed them in the housing box and flushed each one with an equal gas flow rate of 2~l/h. MRPC1a and MRPC1c chambers sandwiched between them the MRPC1b prototype, as it is shown in Fig.~\ref{f5}. The housing box was also flushed with the working gas mixture. 
For each chamber, 16 strips were readout at both ends by NINO based FEE. The LVDS signals delivered by the FEE were fed into the inputs of CAEN V1290A TDCs. The DAQ system was triggered by the coincident signal between MRPC1a and MRPC1c chambers.
\begin{figure}[htb]
\centering
\includegraphics*[width=47mm]{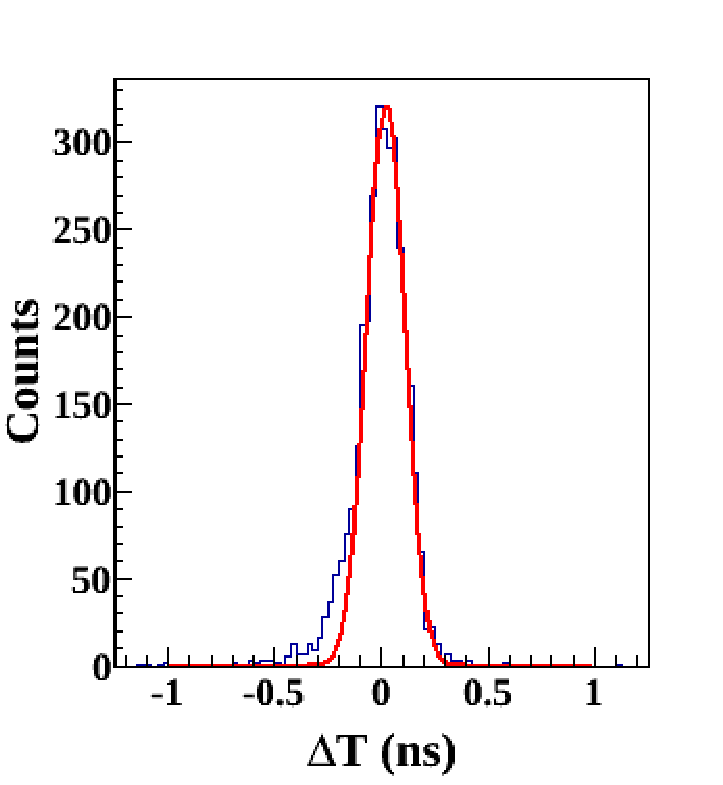}
\caption{Time difference distribution after slweing corrections.}
\label{f5b}
\end{figure}
 The measured efficiency for the MRPC1b chamber, defined as the number of coincident signals in all three chambers divided by the number of coincident signals in MRPC1a and MRPC1c was of 95\% for 2~$\times$~6~kV applied high voltage.  The time difference distribution between MRPC1b and MRPC1a, corrected for the slewing effect, is shown in Fig.~\ref{f5b}. 
\begin{figure}[htb]
\centering
\includegraphics*[width=43.5mm]{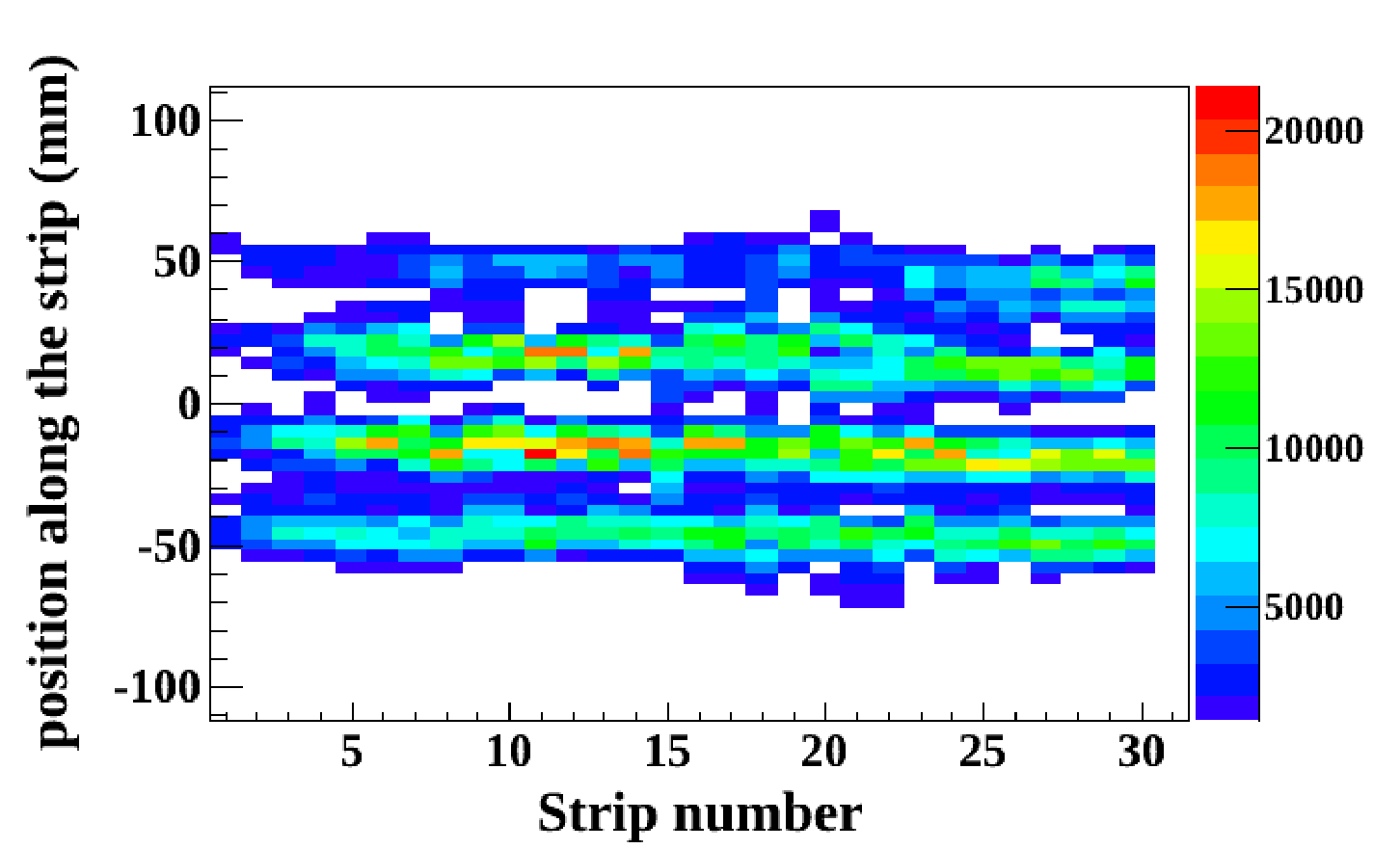}
\includegraphics*[width=43.5mm]{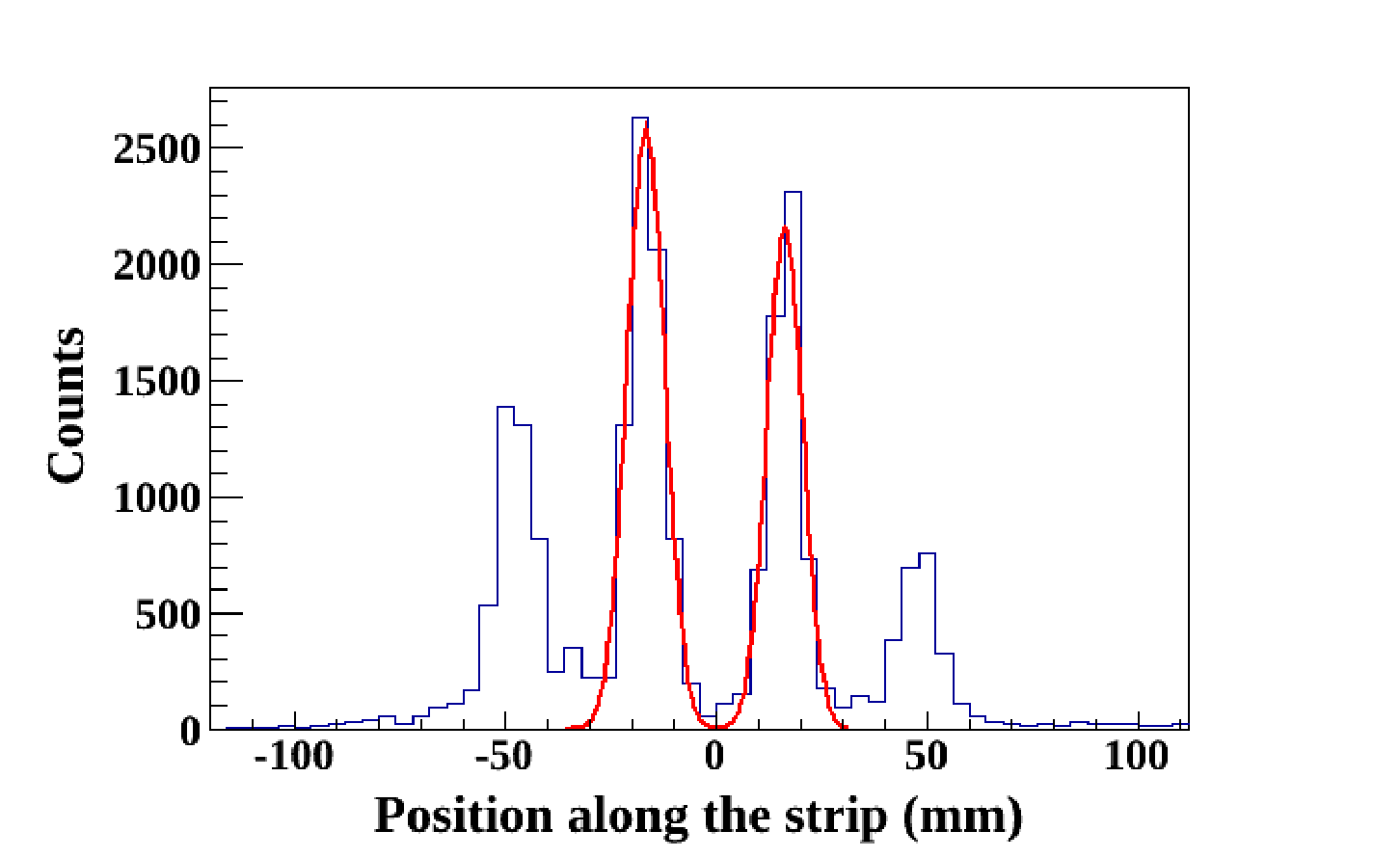}
\caption{Position along the strip versus strip number (left) and projection along one strip (right) for MRPCb.  }
\label{f6}
\end{figure} 
 A single counter time resolution of 62$\pm$2~ps was obtained from the Gauss fit of the time difference spectrum, supposing equal contributions of both chambers.
 
A mapping of the active area of the chambers in a self trigger mode operation of the DAQ system  revealed an activity around the spacer positions. This can be  seen in the two~-~dimensional representation of the position along the strip versus strip number shown in Fig.~\ref{f6}~-~left. The position along the strip was obtained based on the time difference between the two ends of each strip. From the Gauss fit of the two middle peaks seen in the one dimensional projection along one strip (Fig.~\ref{f6}~-~right), an accuracy of the position determination along the strip of 4.5$\pm$0.05~mm was estimated. Based on the distance  between the two middle peaks and the TDC calibration of 25~ps/bin, a  17~cm/ns signal velocity propagation along the strip was determined.  
The mapping in self triggered mode for the MRPC1c prototype evidenced a similar structure with a higher activity around the spacers. Similar results in terms of accuracy of the position determination (4.9$\pm$0.05~mm) and signal propagation velocity (16.5~cm/ns) were obtained.   

\section{Direct flow  MSMGRPC based on discrete spacers}  
\label{3} 
\subsection{Rectangular shape polyimide spacers}
\begin{figure}[htb]
\centering
\includegraphics*[width=50mm]{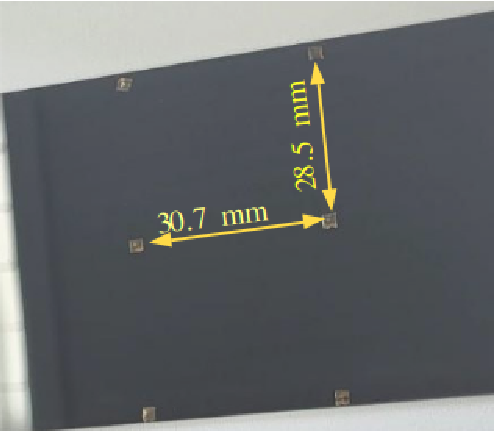}
\caption{Resistive electrode with rectangular shape (2~mm x2~mm) spacers on its surface. } 
\label{f7}
\end{figure}

In order to minimize the aging effects around the spacers, we reduced their size, replacing the continuous nylon fishing line by discrete pad spacers, shown in Fig.~\ref{f7}.  In a first step,  we manufactured rectangular spacers of 2~mm~x~2~mm size and 200~$\mu$m thickness, made of sticky polyimide (kapton type).

  We assembled a direct gas flow chamber with the same inner architecture as previous ones and 56~mm strip length   (MRPC1a described in \cite{nimrpc2022}).  The chamber was equipped with the same type of electronics and operated at 2~$\times$~6~kV. 
\begin{figure}[htb]
\centering
\includegraphics*[width=43mm]{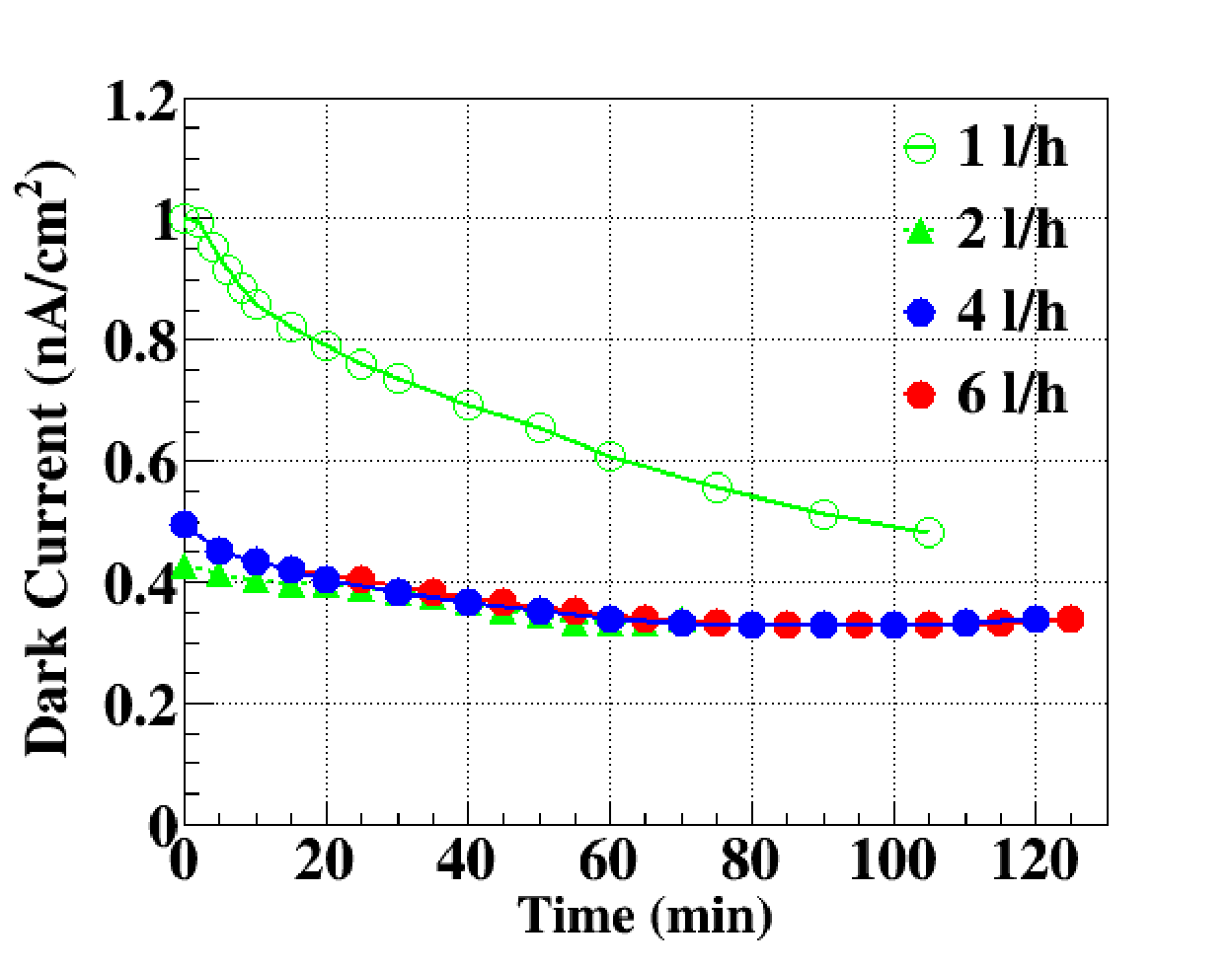}
\includegraphics*[width=43mm]{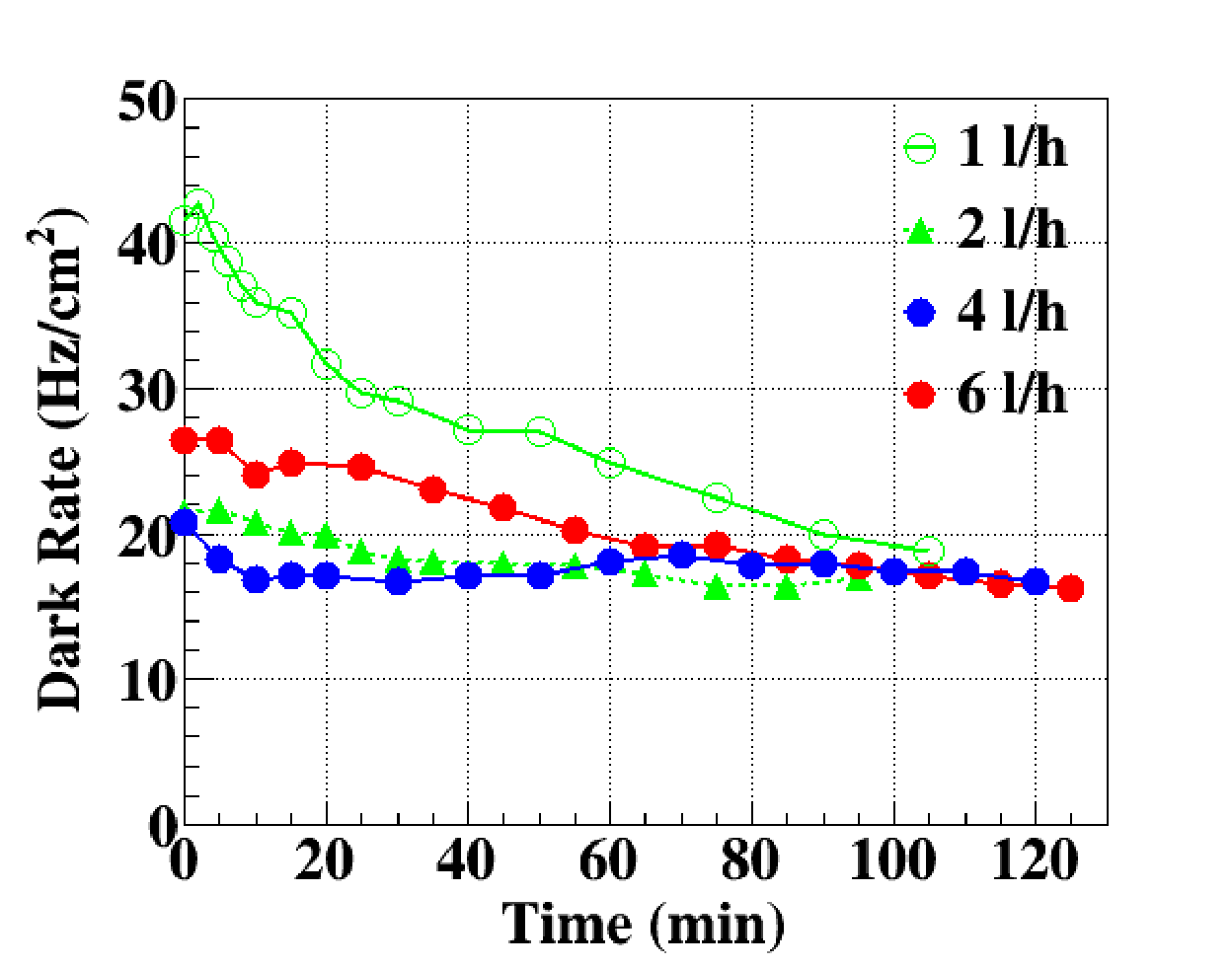}
\caption{Dark current and dark counting rate at different gas flow rates. }
\label{f8}
\end{figure}
It was exposed to high X-ray fluxes corresponding to a maximum counting rate of 29~kHz/cm$^2$, in a similar exposure sequence as described in the section~\ref{aging}. The obtained results showed lower values of the dark current and dark counting rate,  as well as a  much less impact of the gas flow rate (varied between 1~l/h to 6~l/h) on these parameters, as can be seen in Fig.~\ref{f8}. 
Between 2~l/h and 6~l/h gas flow rate it was not observed any difference, the recovery of the dark current and dark counting rate being fast and converging towards similar values in a short time. Based on these positive results, we decreased the flow up to 1 l/h. The measurements showed  still  low values of the dark current and dark counting rate with the same fast recovery in time. 

The two-dimensional mapping of the active area  in self-triggered mode showed that the measured  noisy signals are localized around the positions of the spacers. One possible cause could be that their edges are not so smooth as resulted from the cutting procedure,  having still micro-structure.   
\subsection{Disc shape polyimide spacers}
\begin{figure}[htb]
\centering
\includegraphics*[width=45mm]{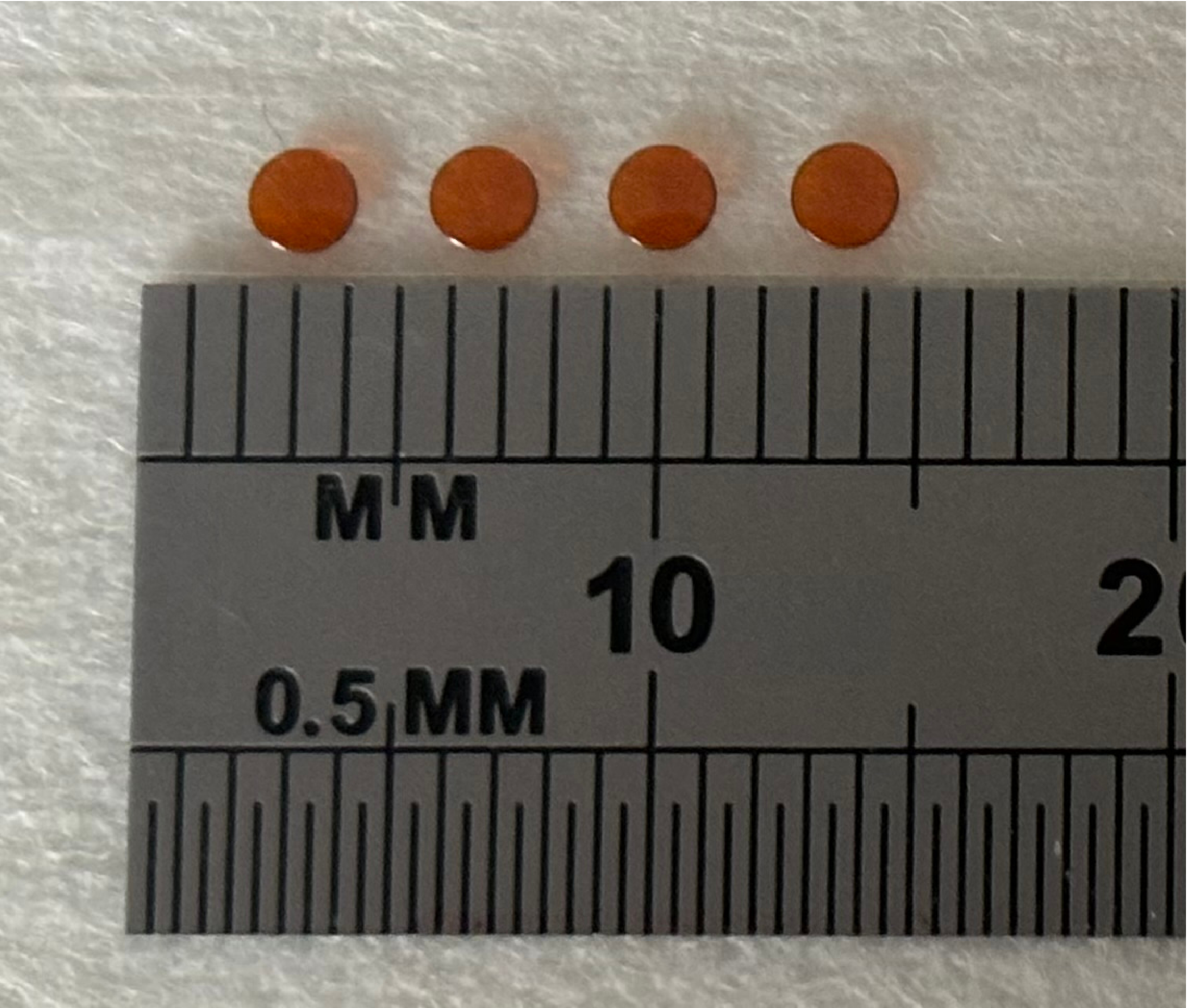}
\caption{Polyimide disc shape spacers.}
\label{f9}
\end{figure}

We found commercially available disk shape spacers of 2~mm diameter, made of polyimide (kapton type), shown in Fig.\ref{f9}, with a maximum available thickness of 170~$\mu$m.
Due to the thinner gas gaps, the new assembled chamber was powered at 2~$\times$~5.7~kV during the aging tests.   
\begin{figure}[htb]
\centering
\includegraphics*[width=50mm]{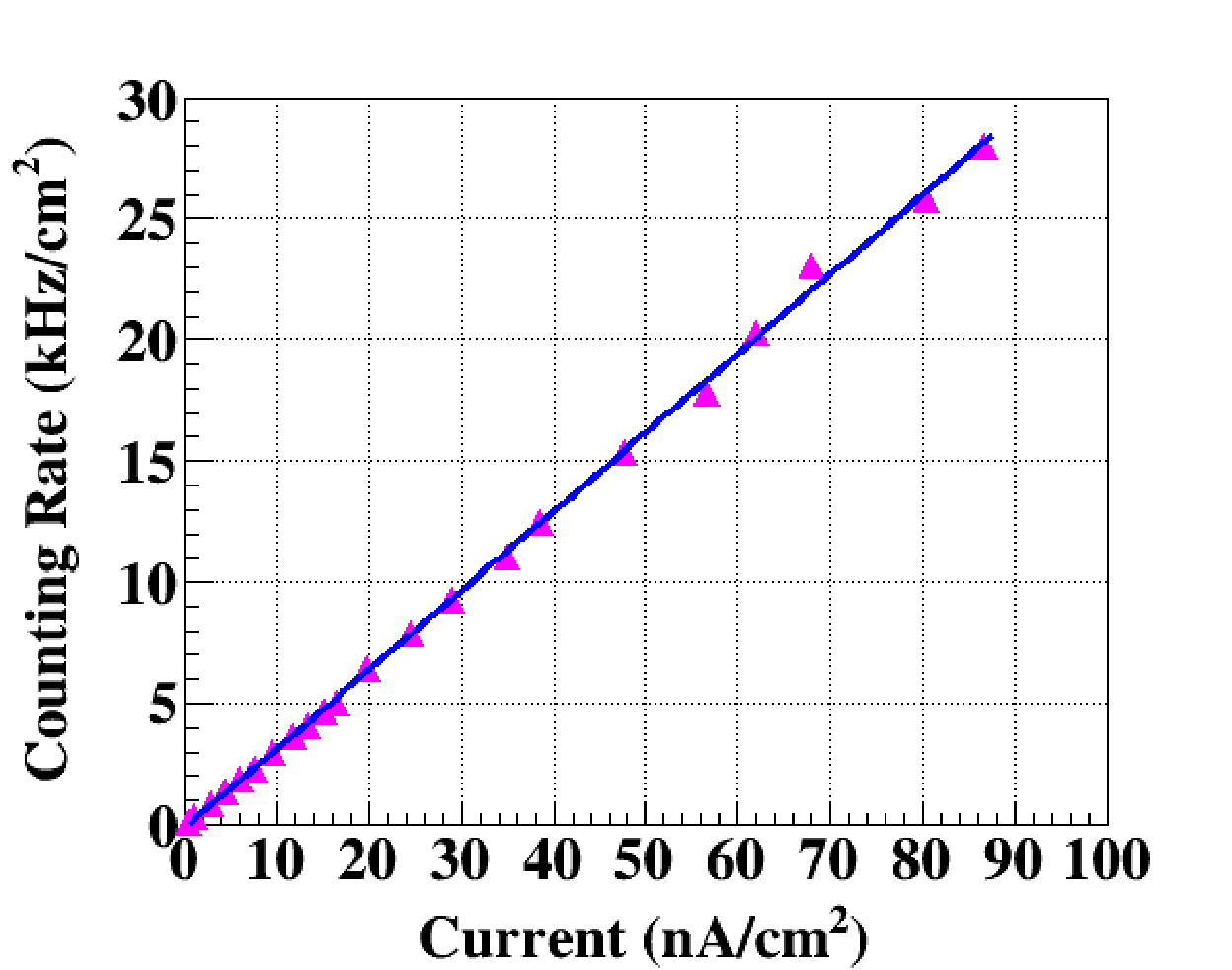}
\caption{ Counting rate versus current in exposure to incresing X-ray flux.}
\label{f10}
\end{figure}
Fig.\ref{f10} shows the measured counting rate versus current for increasing fluxes of X-rays, with a very good linearity up to the maximum measured counting rate of 28~kHz/cm$^2$. 

\begin{figure}[htb]
\centering
\includegraphics*[width=43mm]{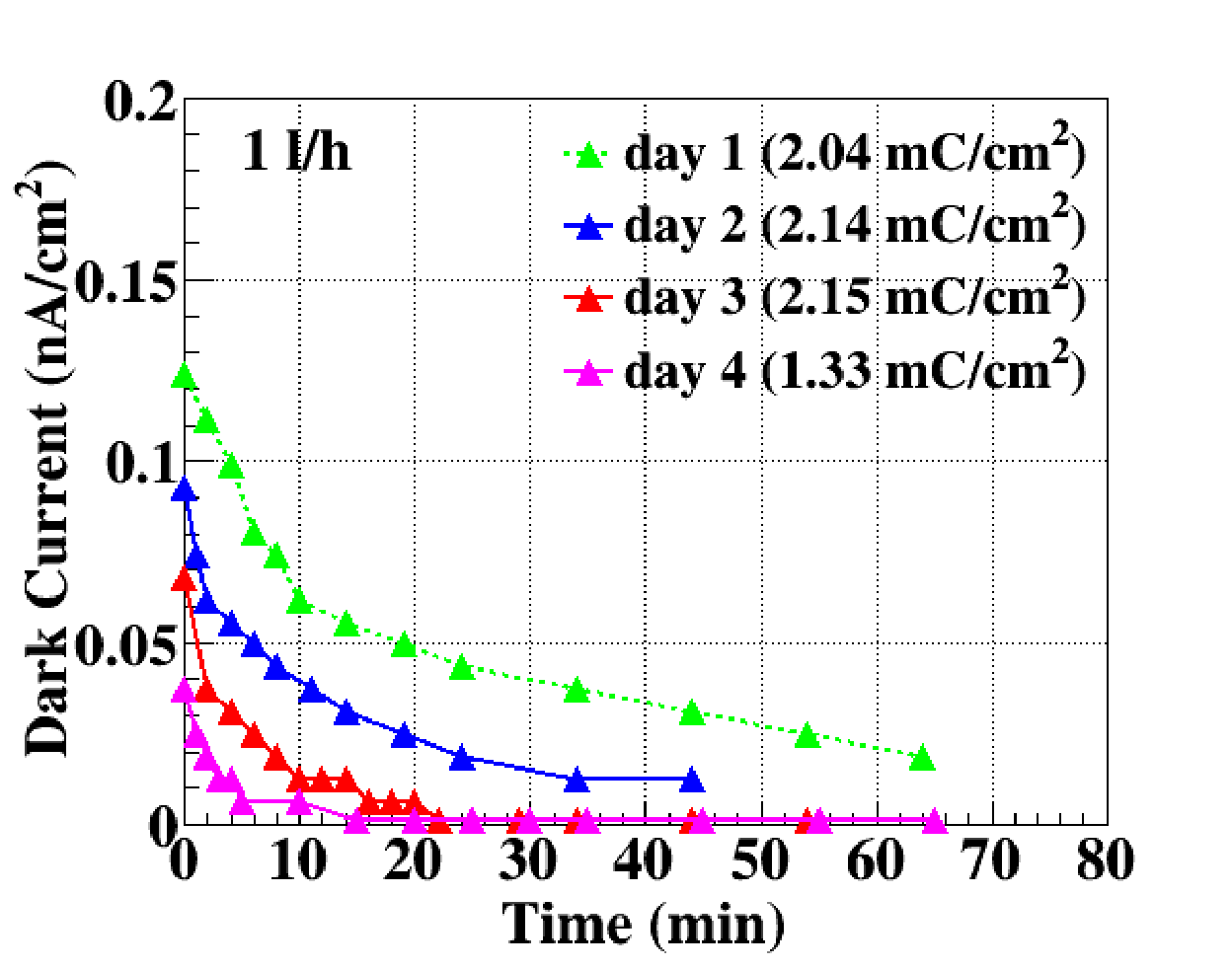}
\includegraphics*[width=43mm]{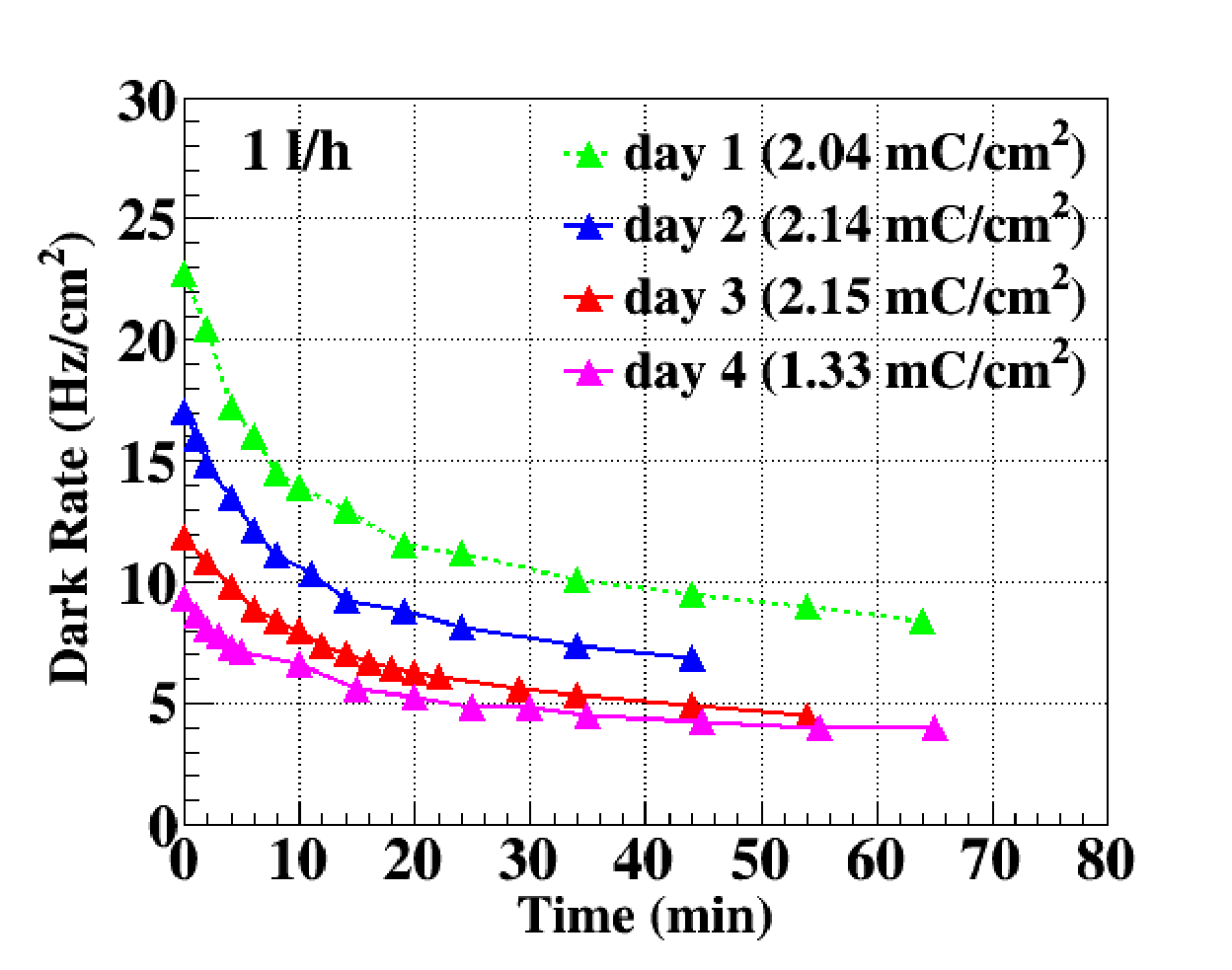}
\caption{Dark current  and dark counting rate after successive exposures. }
\label{f11}
\end{figure}

 Further on, for aging effect investigations, it was exposed on successive days for 6 hours each day, measuring the dark current and dark counting rate recovery in time after each exposure for 2~h, at 1~l/h gas flow rate. As can be seen in Fig.~\ref{f11}, the obtained results showed very low values of the dark current and dark counting rate after each exposure. Moreover, they indicate a conditioning effect which lowered the measured values day by day, as it is shown in Fig.~\ref{f11}. 
We reduced the gas flow rate to 0.5~l/h and exposed the chamber for the same time interval for one more day.  
\begin{figure}[htb]
\centering
\includegraphics*[width=43mm]{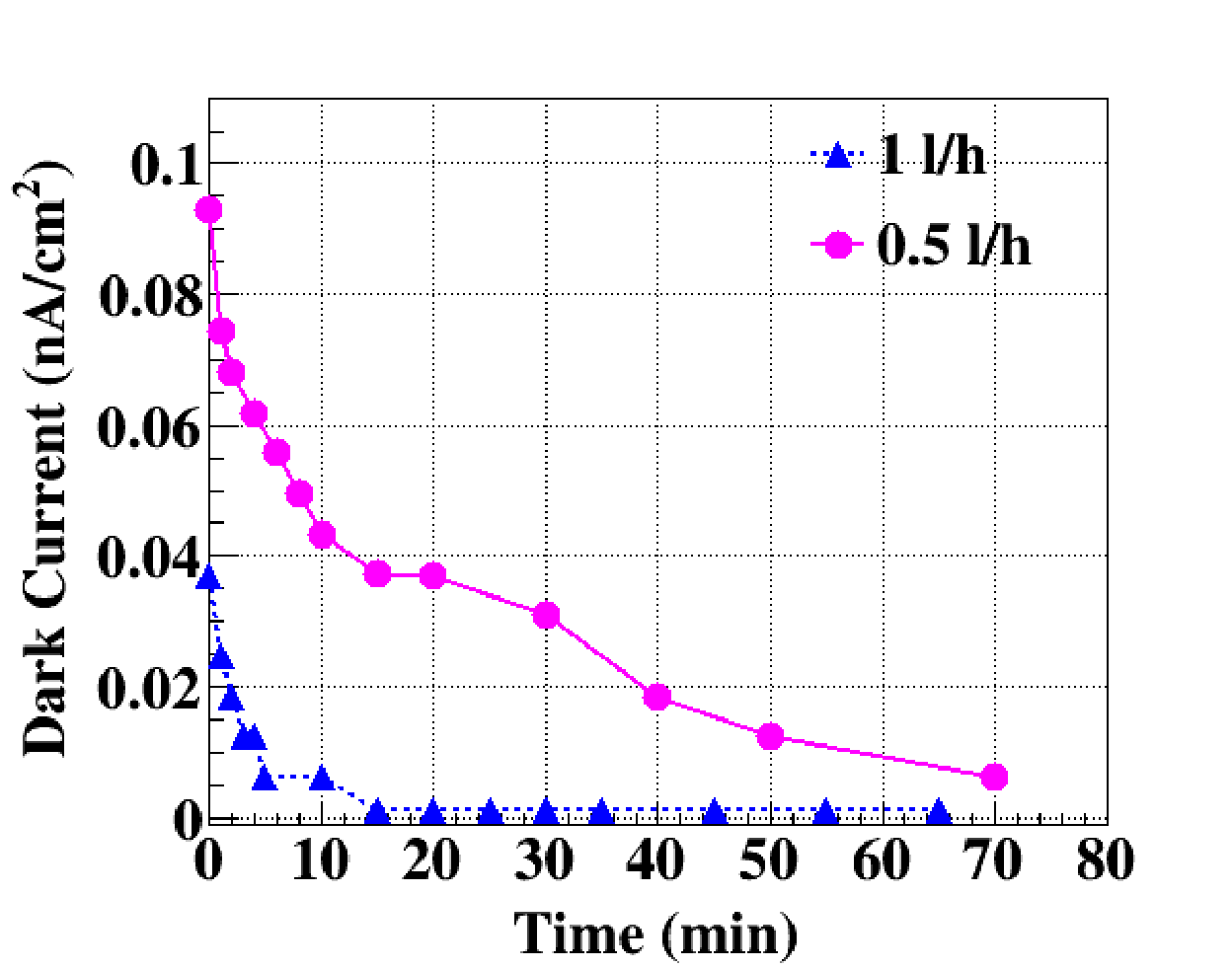}
\includegraphics*[width=43mm]{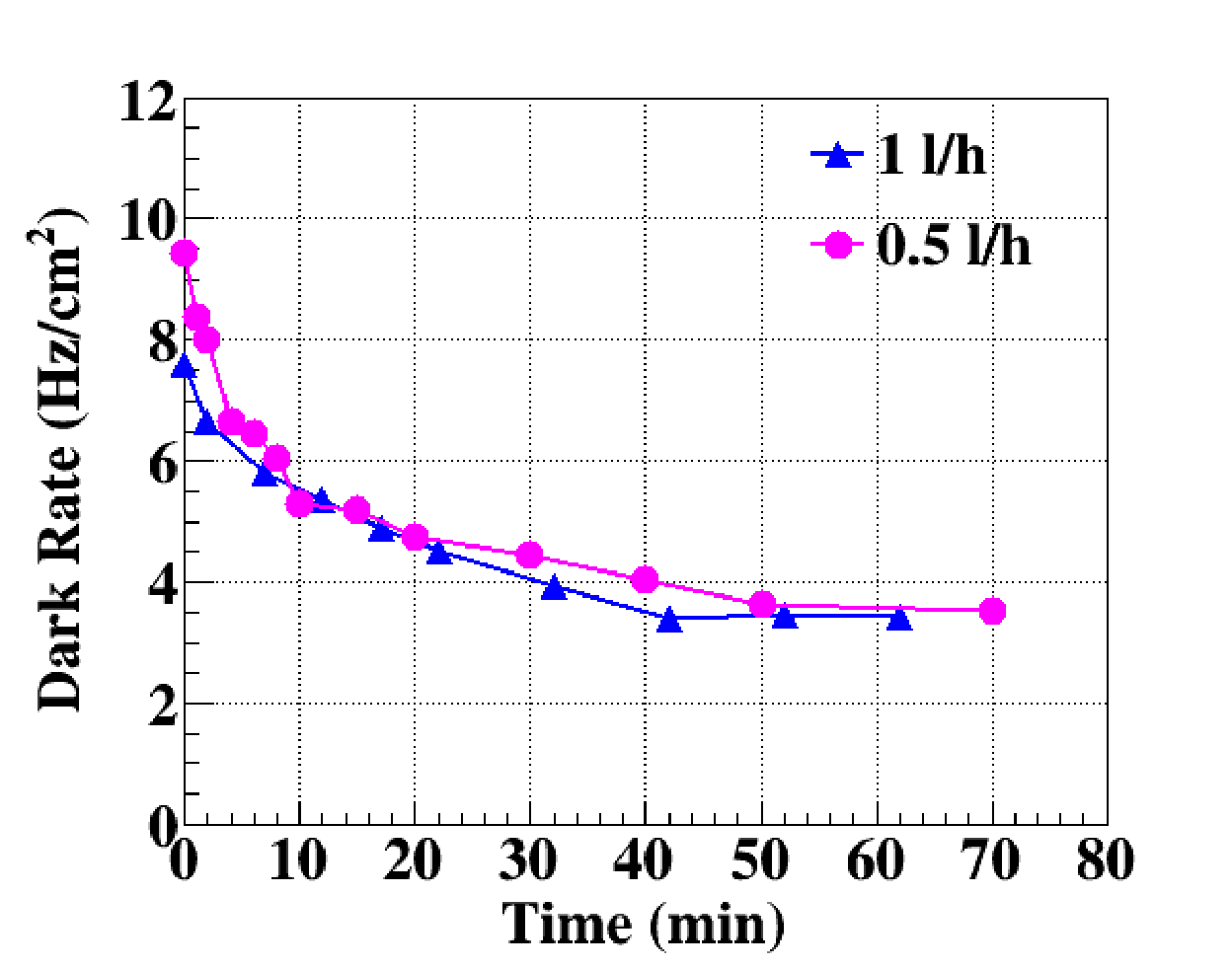}
\caption{Dark current and dark counting rate at 1~l/h and 0.5~l/h gas flow rate. }
\label{f12}
\end{figure}
The results presented in Fig.\ref{f12} show that the dark counting rate has very low values, without significant differences from the ones measured at 1 l/h gas flow rate.  
\section{Conclusions}
\label{4}
The mitigation solutions of the aging effects observed in the MSMGRPC based on gas exchange via diffusion was to modify their architecture in such a way to constrain a  direct flow of the gas mixture through the gaps and to reduce the number of the fishing line spacers inside the active area.
 
 In order to reduce the dark current and the dark counting rate localized around the spacers, a new MSMGRPC architecture based on discrete spacers has been developed, minimizing the area in contact with the spacers. The aging tests with high X-ray fluxes showed a much less dependence of the dark current and counting rate on the gas flow rate,  their values becoming negligible.

Consequently, a direct flow architecture based on discrete spacers is considered as solution for the suppression of the aging effects in MSMGRPCs used for long time periods in high counting rate and high multiplicity environments. However, dedicated high counting rate tests using minimum ionization particles are required in order to confirm the performance of this architecture in real experimental conditions.

\section*{Acknowledgements}
This work was supported by Romanian Government Ministry of Research Innovation and Digitization, projects RO-FAIR 03/16.11.2020 and PN 23210103.


\bibliography{mybibfile}

\end{document}